\documentclass{PoS}
\usepackage{amsmath}

\title{Heavy baryon mass spectrum from lattice QCD with 2+1 flavors}

\ShortTitle{Heavy baryon mass spectrum}

\author{\speaker{Heechang Na} and Steven Gottlieb\\
        Department of Physics, Indiana University, Bloomington, Indiana 47405, USA\\
        E-mail: \email{heena@indiana.edu},
        \email{sg@indiana.edu}}

\abstract{We study the heavy baryon mass spectrum on gauge configurations that
include 2+1 flavors of dynamical improved staggered quarks.
A valence clover heavy quark is combined with two
improved staggered light quark propagators to form baryons with different
flavors.  We are using MILC coarse gauge configurations with a lattice
spacing of about 0.12 fm.
In this preliminary investigation, we explore the chiral limit by
studying two light sea quark masses, three different strange valence quark masses, and 
nine different light valence 
quark masses ranging from 0.1 to 0.4 times the nominal strange quark mass.
}

\FullConference{XXIVth International Symposium on Lattice Field Theory\\
                July 23-28, 2006\\
                Tucson, Arizona, USA}

\begin{document}

\section{Introduction}
Heavy baryons have been extensively investigated by experimental and 
theoretical approaches. From experiment, the singly charmed heavy baryon 
mass spectrum is well known; however, the other heavy baryon masses are 
only crudely known.  From lattice QCD, there are several quenched 
calculations for the heavy baryon mass spectrum [1--6],
and most results are in fair agreement with observed values. 
In this project, we apply 
lattice QCD with dynamical sea quarks for the same objective. 
In the future, we hope to study semi-leptonic decays of heavy baryons \cite{bowler} \cite{tamhankar}.

We present preliminary results for the mass spectrum of five different singly charmed heavy 
 $\frac{1}{2}^+$ baryons: 
$\Lambda_c$, $\Sigma_c$, $\Xi_c$, $\Xi_c^\prime$, and $\Omega_c$. 
In this work, we use two different interpolating operators studied in 
Ref.~\cite{UKQCD} and construct two-point functions using the 
method of Wingate \emph{et al}. \cite{wingate} to combine staggered 
propagators for the light valence quarks
and a Wilson type (clover) propagator for the heavy valence quark.

\section{Construction of two-point functions}
The interpolating operators 
to describe $\frac{1}{2}^+$ singly heavy baryons are 
\cite{UKQCD} 
\begin{equation}
\mathcal{O}_5 = \epsilon_{abc} ( \psi_1^{aT} C \gamma_5 \psi_2^b ) \Psi_H^c , \;\;\;\;\;\;\;
\mathcal{O}_\mu = \epsilon_{abc} ( \psi_1^{aT} C \gamma_\mu \psi_2^b ) \Psi_H^c,
\label{operator}
\end{equation}
where $\epsilon_{abc}$ is the Levi-Civita tensor, $\psi_1$
and $\psi_2$ are light valence quark fields for up, down, or strange quarks,
$\Psi_H$ is the heavy valence quark field, 
and $C$ is the charge conjugation matrix. 
As can be seen, the spinor indices of the light quark fields
are contracted together, so the spinor index of the operator comes from 
$\Psi_H$.
Basically, $\mathcal{O}_5$ is the operator for 
$s^\pi = 0^+$, and  $\mathcal{O}_\mu$ is for
$s^\pi = 1^+$, where $s^\pi$ is spin parity state for light quark pair. Each heavy baryon is obtained by choosing different quark flavor combinations,
shown in Table~\ref{baryonT}.

\begin{table}
\centering
\begin{tabular}{|c|c|c|c|c|}
\hline
Baryon&Content&Operator&$J^p$&$s^\pi$\\
\hline
$\Lambda_c$&$llc$&$\mathcal{O}_5$&&$0^+$\\
$\Xi_c$&$lsc$&$\mathcal{O}_5$&&$0^+$\\
$\Sigma_c$&$llc$&$\mathcal{O}_\mu$&$\frac{1}{2}^+$&$1^+$\\
$\Xi_c^\prime$&$lsc$&$\mathcal{O}_\mu$&&$1^+$\\
$\Omega_c$&$ssc$&$\mathcal{O}_\mu$&&$1^+$\\
\hline
\end{tabular}
\caption{Heavy baryons, flavor contents and operators. $J^p$ is total spin parity
 of the heavy baryon, and $s^\pi$ is spin parity of the light quark pair. In the quark content column, 
 $l$ is the light valence quark, $s$ is the strange quark, and $c$ is the charm 
 quark.}
\label{baryonT}
\end{table}

Since we use staggered fermion propagators for the light quarks, and a Wilson type propagator for
the heavy quark, 
we will use the method of Wingate \emph{et al}. \cite{wingate} to convert from a staggered 
propagator to a naive quark propagator. They show
\begin{equation}
G_\psi(x,y)=\Omega(x)\Omega^\dagger(y)G_\chi(x,y)
\label{win1}
\end{equation}
where,
\begin{equation}
\Omega(x)=\prod_\mu(\gamma_\mu)^{x_\mu/a}.
\end{equation}
$G_\psi(x,y)$ is the naive quark propagator, and $G_\chi(x,y)$ is the 
staggered quark propagator.
With this basic relationship, we can construct the heavy baryon 
two-point functions 
\begin{eqnarray}
C_{\alpha\beta}(\vec{p},t)&=&\sum_{\vec{x}}e^{-i \vec{p} \cdot \vec{x}} \langle 
\mathcal{O}_{5\alpha} (\vec{x},t) \overline{\mathcal{O}}_{5\beta} (\vec{0},0) \rangle \\
&=&\sum_{\vec{x}}e^{-i\vec{p}\cdot\vec{x}} \epsilon_{abc} \epsilon_{a'b'c'} \mathbf{tr} [ 
G_1^{aa' T} (x,0) C \gamma_5 G_2^{bb'} (x,0) C \gamma_5 ] G_{H\alpha\beta}^{cc'} (x,0).
\label{trace}
\end{eqnarray}
$G_1(x,0)$ and $G_2(x,0)$ are naive light quark propagators, $G_H(x,0)$ is the heavy quark
propagator, and $\alpha$ and $\beta$ are spinor indices. As expected from the property of the interpolating 
operator, the spinor structure of the heavy baryon is determined from the heavy quark. Note that the trace in
Eq.~\ref{trace} is for spinor indices, not for color indices. Now, using Eq.~\ref{win1},
\begin{equation}
 G_2^{bb'} (x,0) = \Omega(x) \Omega^\dagger(0) G_{2\chi}^{bb'} (x,0) = 
 \Omega(x) G_{2\chi}^{bb'} (x,0)
\end{equation}
and
\begin{equation}
G_1^{aa'T} (x,0) = \Omega^T(x) G_{1\chi}^{aa'} (x,0).
\end{equation}
So
\begin{equation}
C_{\alpha\beta}(\vec{p},t)=\sum_{\vec{x}}e^{-i\vec{p}\cdot\vec{x}} \epsilon_{abc} \epsilon_{a'b'c'}
 \mathbf{tr} [ \Omega^T(x) C \gamma_5 \Omega(x) C \gamma_5 ]  G_{1\chi}^{aa'} (x,0)
 G_{2\chi}^{bb'} (x,0) G_{H \alpha\beta}^{cc'} (x,0).
\end{equation}
But the trace can be simplified:
\begin{equation}
\mathbf{tr} [ \Omega^T(x) C \gamma_5 \Omega(x) C \gamma_5 ]=
\mathbf{tr} [ -1(-1)^{x_1+x_3}(-1)^{x_1+x_3} ]=-4.
\end{equation}
Finally, the two-point function can be expressed as,
\begin{equation}
C_{\alpha\beta}(\vec{p},t)=\sum_{\vec{x}}e^{-i\vec{p}\cdot\vec{x}} \epsilon_{abc} \epsilon_{a'b'c'}
(-4) G_{1\chi}^{aa'} (x,0) G_{2\chi}^{bb'} (x,0) G_{H \alpha\beta}^{cc'} (x,0).
\end{equation}
This two-point function is constructed using the operator $\mathcal{O}_5$. Similarly, we can derive the two-point function with the other operator $\mathcal{O}_\mu$.

\section{Data analysis}

We use two ensembles of  $20^3\times64$ MILC coarse dynamical lattice gauge configurations 
with lattice spacing $a\approx 0.12$ fm. We investigate 385 configurations for the ensemble with
$am_l=0.01$, and $am_s=0.05$, and 458 configurations for the ensemble with
$am_l=0.02$, and $am_s=0.05$,  where $m_l$ is the light sea quark mass, 
and $m_s$ is the strange sea quark mass. 
For each configuration, we require several propagators for the valence quarks. 
We compute nine different staggered light quark propagators with masses between
0.005 and 0.02, and three staggered strange quark propagators with masses 0.024, 0.03, 
and 0.0415. For the heavy charm quark, we use only one kappa value $\kappa=0.122$
based on tuning for our heavy-light meson decay constant calculation \cite{bare}.

For fitting the baryon propagators, we use correlated least squares fits, and for error
estimation, we generate 1000 bootstrap samples. 
Taking into account the periodic boundary condition in time for the valence quarks and the staggered phases, the fit model is 
\begin{align} 
P(t)= &A e^{-mt} + A e^{-m(T-t)} +(-1)^t \tilde{A} e^{- \tilde{m} t} + (-1)^t 
\tilde{A} e^{- \tilde{m} (T-t)} \notag \\
&+ A_* e^{-m_* t} + A_* e^{-m_* (T-t)} +(-1)^t \tilde{A}_* e^{- \tilde{m}_* t} + (-1)^t
\tilde{A}_* e^{- \tilde{m}_* (T-t)},
\end{align}
where $m$ is the ground state of positive parity, $\tilde{m}$ has opposite 
parity, and the values with an asterisk denote the corresponding excited states.
In practice, most propagators could be fit including contributions with 
just two or three particles. 
In most cases the confidence level of the fit is between 40 and 60 percent. 
(Of course, on each ensemble, baryon propagators with different valence 
masses are correlated.)

\begin{figure}[ht]
\centering
\includegraphics[width=.45\textwidth]{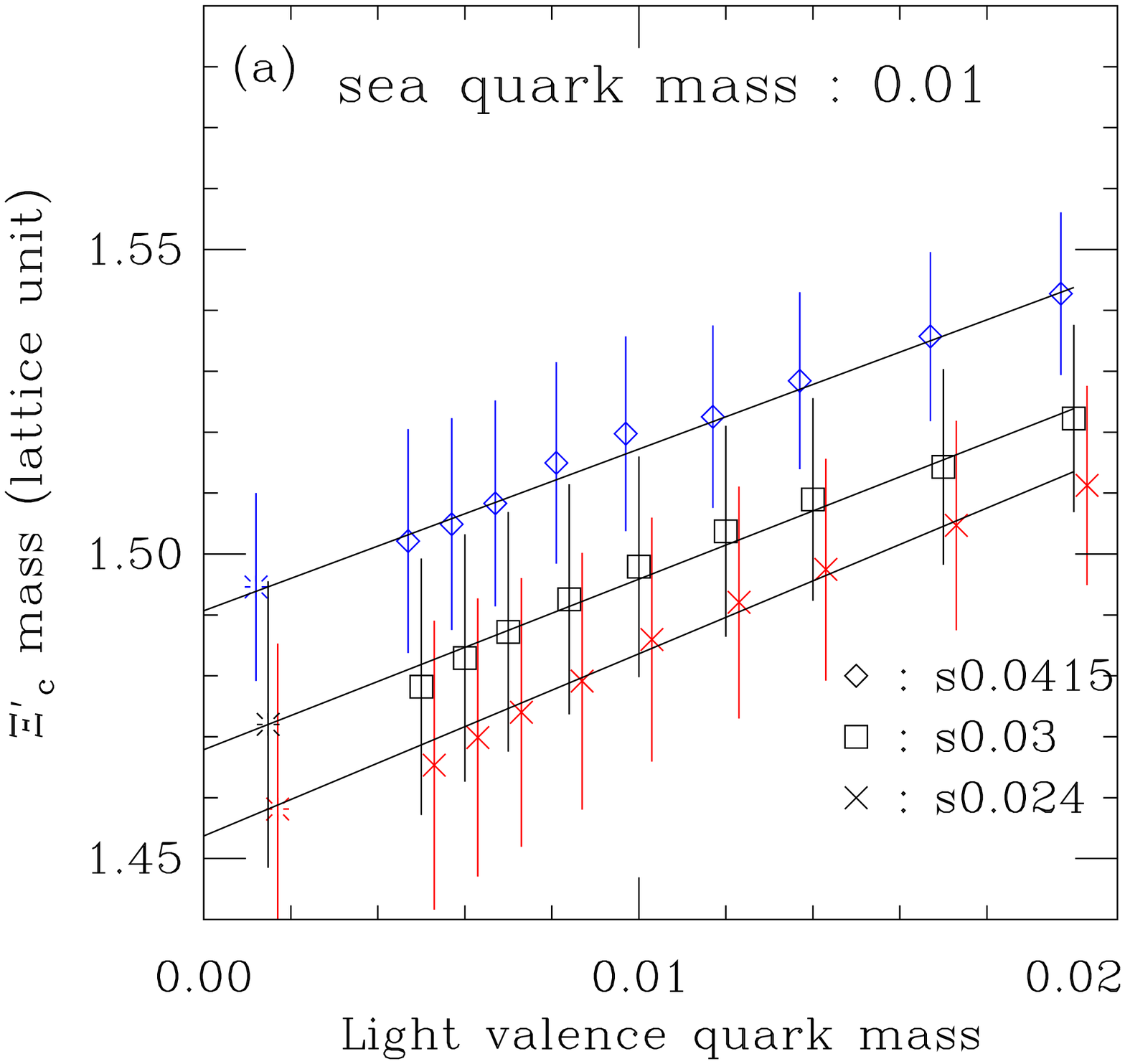}
\includegraphics[width=.45\textwidth]{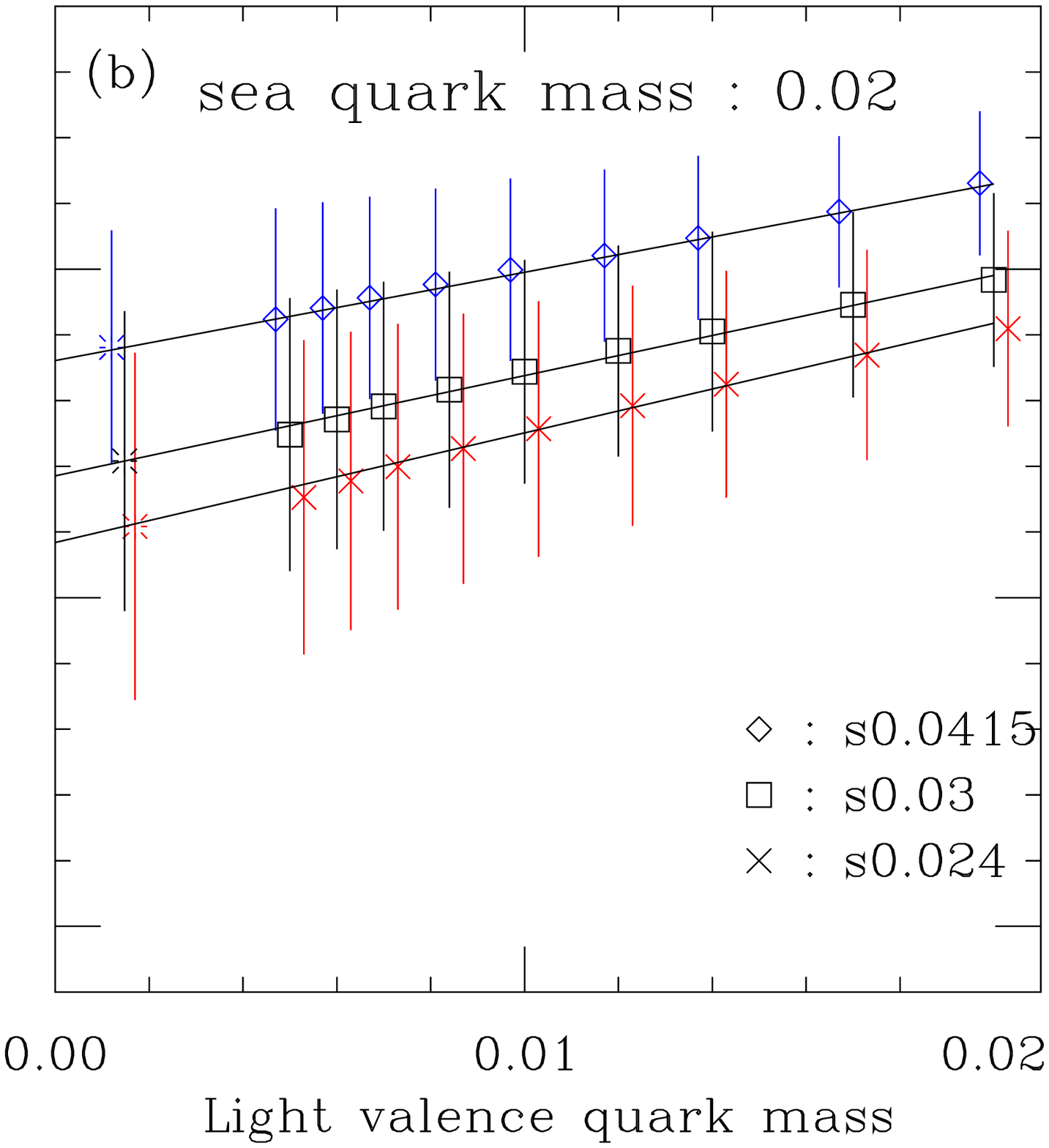}
\caption{Chiral extrapolation of light valence quark mass on each ensemble.}
\label{fig1}
\end{figure}

\begin{figure}
\centering
\includegraphics[width=.48\textwidth]{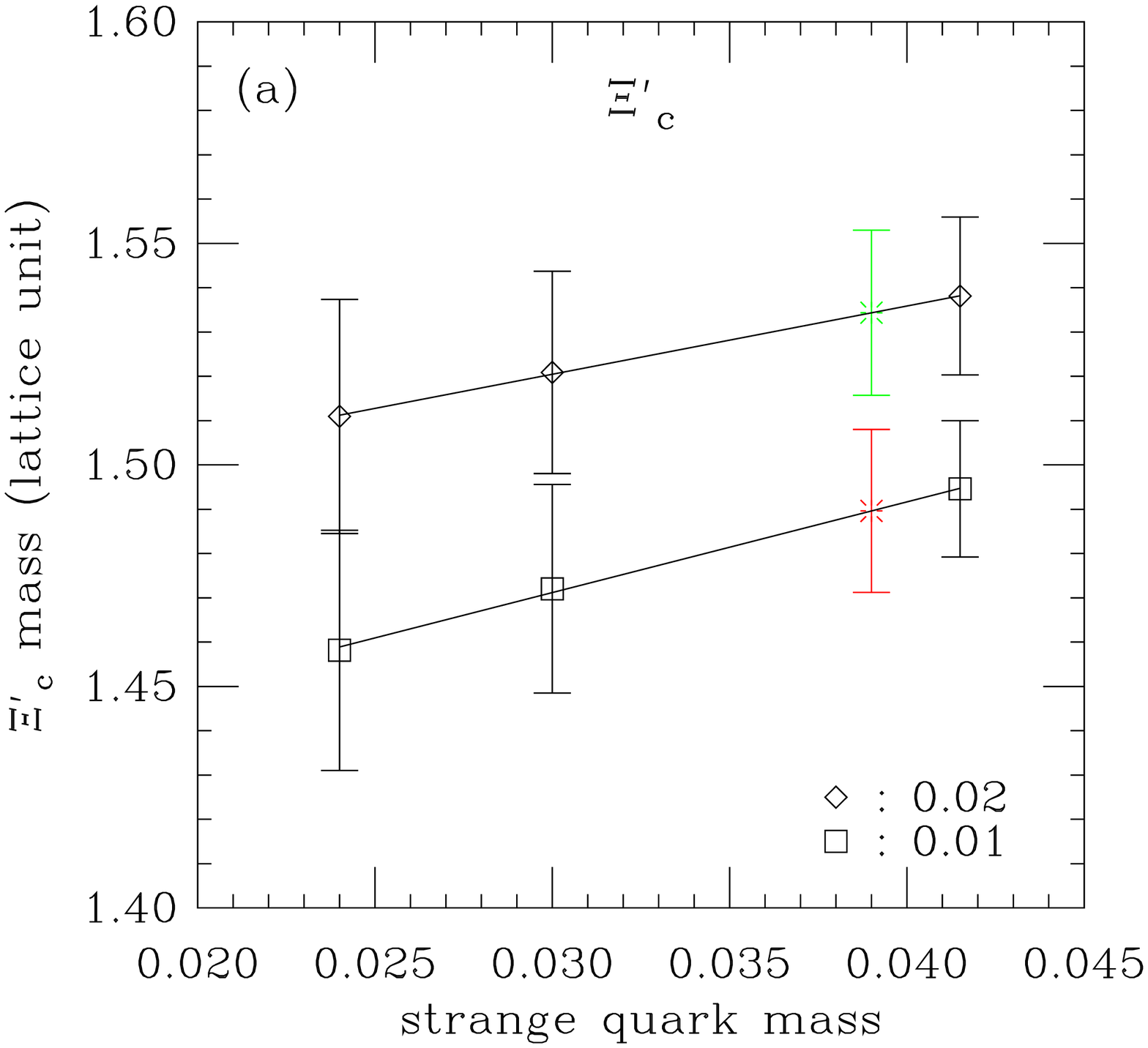}
\includegraphics[width=.48\textwidth]{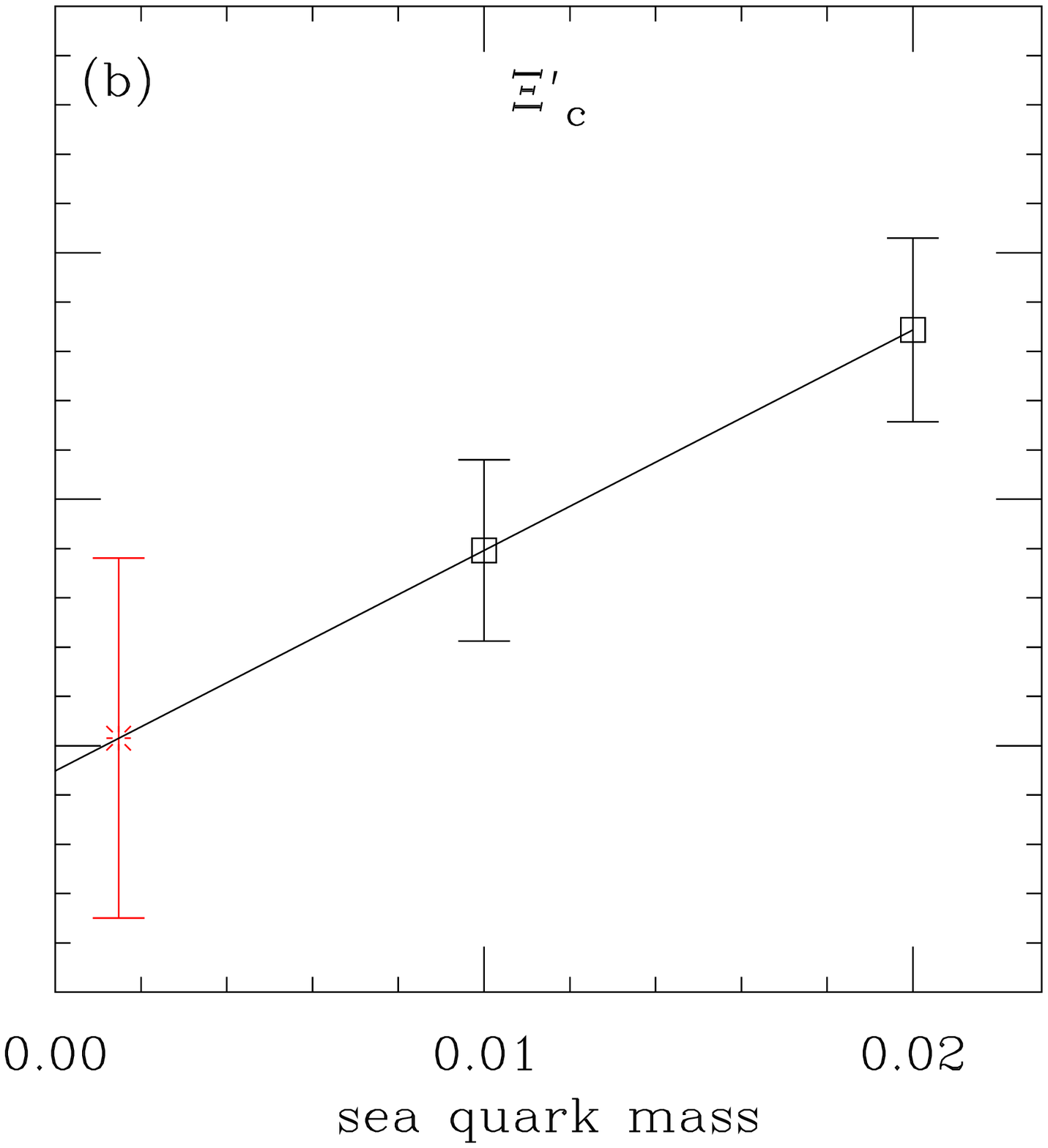}
\caption{Interpolation of strange quark mass (a), and extrapolation of light sea quark mass
(b). 
The bare strange quark mass is 0.039, and the bare light sea quark mass is 0.00148.
Bursts are the interpolated and extrapolated results.}
\label{fig2}
\end{figure}

To discuss extrapolation to the chiral limit, we present an example for 
$\Xi_c^\prime$. First, let's look at the extrapolation of the light 
valence quark mass shown in Fig.~\ref{fig1}. 
As we can see, there are three sets of 
data for each ensemble corresponding to each valence strange quark mass.
We have performed linear extrapolations although there is a very small 
curvature seen for the lighter sea quark mass (Fig.~\ref{fig1}(a)). 
The bursts show the linear extrapolation to the bare light quark 
mass 0.00148 \cite{real} in lattice units on this ensemble.
Next, we interpolate in the strange quark mass (Fig.~\ref{fig2}(a)) and 
extrapolate in the light sea quark mass (Fig.~\ref{fig2}(b)). 
The bare strange quark mass was estimated to be 0.039 \cite{real}.
In Fig.~\ref{fig2}(a), we see that linear interpolation in the strange 
quark mass is perfectly adequate. 

\begin{figure}[ht]
\centering
\includegraphics[width=.6\textwidth]{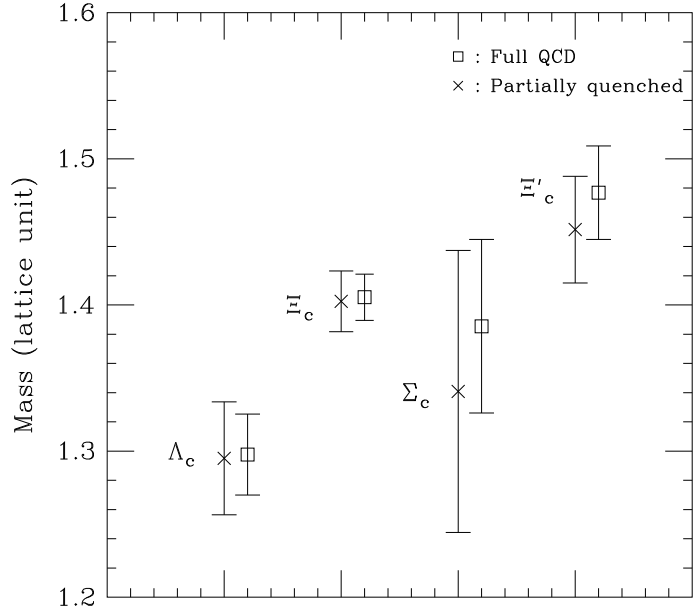}
\caption{Full QCD compared with partially quenched QCD chiral extrapolation. Four different masses of singly heavy baryons.
Note that $\Omega_c$ dose not appear, since $\Omega_c$ does not contain light valence quarks.}
\label{fig3}
\end{figure}

After extrapolation in the light sea quark mass, we find the mass of 
$\Xi_c^\prime$ within the partially quenched approximation.
Alternatively, there is a full QCD point for each ensemble, and we can 
simultaneously extrapolate in light sea and valence masses.
Fig.~\ref{fig3} compares the results of these two approaches for four baryons.
As we can see, the full QCD results have smaller error bars than the partially
quenched results.
Our multistep partially quenched interpolation and extrapolation 
procedure may be causing larger errors than the simple full 
QCD extrapolation in light quark mass.
If we had fit all the data with a partially quenched formula depending 
on both sea and valence
quark masses, the partially quenched QCD chiral extrapolation 
might have given smaller error bars than the fit to the limited full QCD data.
(We plan to try that in the future.)
We present our preliminary results here based upon the full QCD extrapolation.

\section{Results}

\begin{figure}
\centering
\includegraphics[width=.48\textwidth]{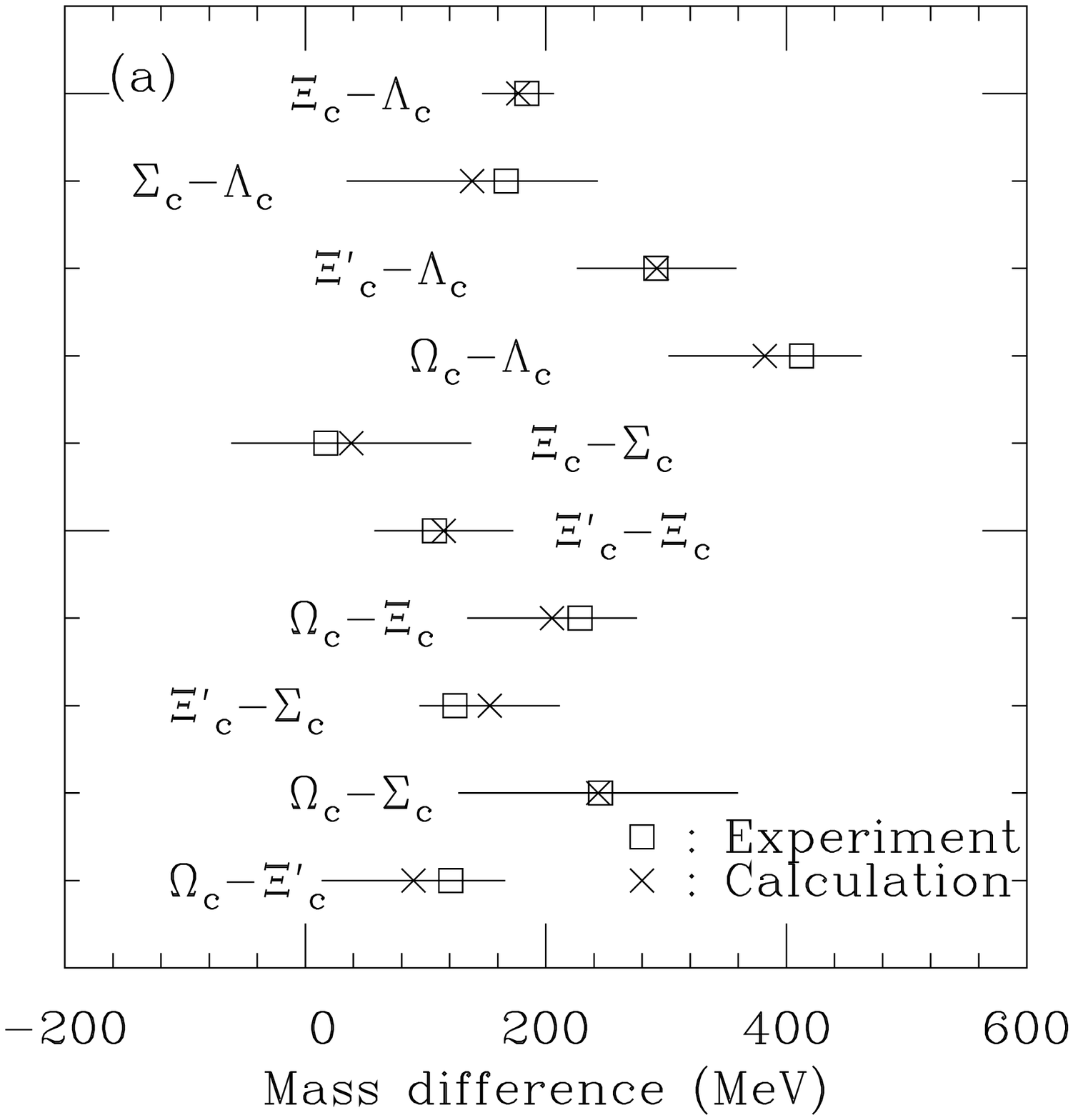}
\includegraphics[width=.48\textwidth]{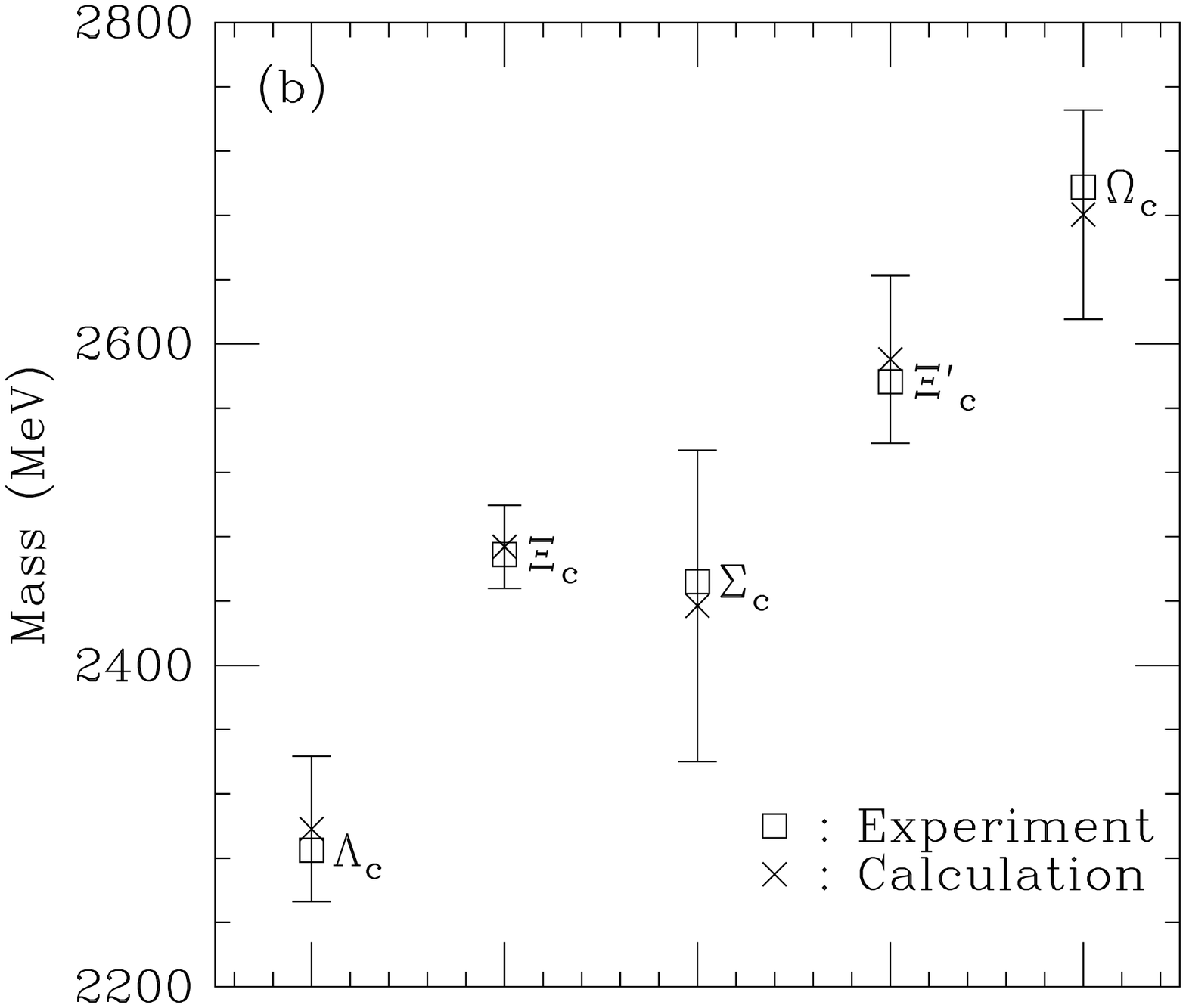}
\caption{Mass splitting (a),
and mass spectrum (b) of five different singly charmed heavy baryons 
with $\frac{1}{2}^+$ in MeV units.}
\label{fig4}
\end{figure}

We present mass splittings in MeV between five different singly heavy
baryons in Fig.~\ref{fig4}(a). Square data points correspond to
the experiment data 
\cite{pdg}, and crosses with error bars indicate our results. The experiment 
errors are considerably smaller than our own, so we ignore them here.
Our results 
agree with the experimental values within errors.
In fact, the agreement is so good, we wonder if we are overestimating our 
errors.  On the other hand, all the baryon masses are correlated.

We also present the mass spectrum itself in Fig.~\ref{fig4}(b). In order to get the 
physical mass $M_{phy}$ of a heavy baryon,  
we need to add a constant mass shift $\Delta$ to the calculated mass 
$M_{cal}$, i.e.
\begin{equation}
M_{phy}=M_{cal}+\Delta.
\end{equation}
In this project, we chose the constant mass shift $\Delta$ in a simple way
to best match the average of the experimental masses. 
\begin{equation}
\Delta = Average(M_{exp}-M_{cal}).
\end{equation}
We should investigate the momentum dependence of the heavy baryon energy
in order to calculate the kinetic mass of the state.
If we use the dispersion relation, we can take the kinetic 
mass $M_{kin}$ as the physical mass $M_{phy}$ \cite{wingate},
\begin{equation}
M_{kin}=\frac{|\vec{p}|^2-[E(\vec{p})-E(0)]^2}{2[E(\vec{p})-E(0)]}.
\end{equation}

\section{Future study}

We have presented preliminary spectrum results based on dynamical lattice QCD. 
The study can be extended in several ways. 
We shall investigate singly bottom baryons and doubly charmed baryons. 
Recently, the SELEX collaboration reported the measurement of the 
doubly charmed baryon $\Xi_{cc}^+$ \cite{selex}, 
so it would be timely to compute the doubly charmed baryon spectrum.
So far we have only studied $\frac{1}{2}^+$ states, 
and we would like to extend the study to $\frac{3}{2}^+$  states. In this way, we can investigate
the hyperfine splitting problem \cite{woloshyn2} \cite{khan} in heavy baryons.
In addition, we can increase statistics by using more configurations and ensembles, and 
additional time sources for our propagators.
We also need to clarify the taste structure of our baryon operators.
\acknowledgments{We thank Arifa Ali Khan for discussions and encouragement,
and Norman Christ for asking about taste mixing.}


\begin{thebibliography}{99}
  \bibitem{UKQCD} K.C.~Bowler \emph{et al}. (UKQCD Collaboration), Phys. Rev. D {\bf 54} (1996) 3619.

  \bibitem{woloshyn1} N.~Mathur \emph{et al}., Phys. Rev. D {\bf 66} (2002) 014502.

  \bibitem{woloshyn2} R.M.~Woloshyn, Phys. Lett. B {\bf 476} (2000) 309.

  \bibitem{tamhankar} S.~Gottlieb and S.~Tamhankar, Nucl. Phys. Proc. Suppl. {\bf 119} 
  (2003) 644.

  \bibitem{khan} A.~Ali Khan \emph{et al}., Phys. Rev. D {\bf 62} (2000) 054505.

  \bibitem{chiu} T.W.~Chiu and T.H.~Hsieh, Nucl. Phys. A {\bf 755} (2005) 471c.

  \bibitem{bowler} K.C.~Bowler \emph{et al}. (UKQCD Collaboration), Phys. Rev. D {\bf 57} (1998) 6948.

  \bibitem{wingate} M.~Wingate \emph{et al}., Phys. Rev. D {\bf 67} (2003) 054505.

  \bibitem{bare} C.~Aubin \emph{et al}. (Fermilab Lattice, MILC, and HPQCD Collaborations), 
  Phys. Rev. Lett. {\bf 95} (2005) 122002.

  \bibitem{real} C.~Aubin \emph{et al}. (MILC Collaboration), Phys. Rev. D {\bf 70}
  (2004) 114501.

  \bibitem{pdg} W.M.~Yao \emph{et al}., J. Phys. G {\bf 33} (2006) 1.

  \bibitem{selex} A.~Ocherashvili \emph{et al}., Phys. Lett. B {\bf 628} (2005) 18.
\end{thebibliography}
\end{document}